\documentclass[12pt]{article}
\pdfoutput=1

\usepackage[style=ieee,citestyle=numeric,doi=false]{biblatex}
\renewbibmacro*{bbx:savehash}{}

\usepackage[english]{babel}
\usepackage{graphicx}
\usepackage{amsmath}
\usepackage{amssymb}
\usepackage{multirow}
\usepackage{url}
\usepackage{tikz}
\usetikzlibrary{decorations.markings}
\usepackage{subcaption}

\allowdisplaybreaks
\def\eq#1{(\ref{#1})}
\def\[#1\]{\begin{align}#1\end{align}}

\begin{document}

\begin{titlepage}
\title{
\hfill\parbox{4cm}{ \normalsize YITP-17-34}\\ 
\vspace{1cm} 
Symmetric configurations highlighted \\
by collective quantum coherence
}
\author{
Dennis Obster$^{1,2}$\footnote{dobster@science.ru.nl}, 
Naoki Sasakura$^2$\footnote{sasakura@yukawa.kyoto-u.ac.jp}
\\
$^1${\small{\it Institute for Mathematics, Astrophysics and Particle Physics, Radboud University,}}
\\ {\small{\it Heyendaalseweg 135, 6525 AJ Nijmegen,The Netherlands}}
\\
$^2${\small{\it Yukawa Institute for Theoretical Physics, Kyoto University,}}
\\ {\small{\it  Kitashirakawa, Sakyo-ku, Kyoto 606-8502, Japan}}
}

\date{\today}

\maketitle

\begin{abstract}
Recent developments in quantum gravity have shown the Lorentzian treatment to be
a fruitful approach towards the emergence of macroscopic spacetimes. 
In this paper, we discuss another related
aspect of the Lorentzian treatment: we argue that collective quantum coherence
may provide a simple mechanism for highlighting symmetric configurations over generic non-symmetric ones.
After presenting the general framework of the mechanism,
we show the phenomenon in some concrete simple examples
in the randomly connected tensor network, which is tightly 
related to a certain model of quantum gravity, i.e., the canonical tensor model.
We find large peaks at configurations invariant under Lie-group symmetries as well as 
a preference for charge quantization, even in the Abelian case.
In future study, this simple mechanism may provide a way to analyze the 
emergence of macroscopic spacetimes with global symmetries
as well as various other symmetries existing in nature, which are usually postulated.
\end{abstract}

\end{titlepage}

\section{Introduction}

In recent years there have been some developments which show that the Lorentzian treatment of quantum gravity 
can be very successful in describing spacetime. One is the success of 
causal dynamical triangulation (CDT) \cite{Ambjorn:2004qm}. 
CDT takes into account a causality condition, distinguishing Lorentzian spacetimes from the Euclidean ones, 
and has been shown to generate macroscopic spacetimes similar to the de~Sitter spacetime, 
which is a good approximation 
of our actual universe. This is in contrast with the Euclidean counterpart, in which 
the emergence of macroscopic spaces seems to be more non-trivial~\cite{Coumbe:2014nea,Laiho:2016nlp}.
Another development is the argument by van Raamsdonk 
that quantum correlations are the essence of connectivity of classical spacetimes,
which has been derived in the context of AdS/CFT correspondence \cite{VanRaamsdonk:2010pw}.
This seems to imply that the quantum correlations 
of a model for quantum gravity are crucial for the emergence of macroscopic spacetimes.
It was also argued that Lorentzian path integral has some advantages over the Euclidean one 
in describing the wave function of universe in the mini-superspace approximation \cite{Feldbrugge:2017kzv}.  

In this paper, we discuss another related aspect of Lorentzian treatment in quantum gravity:   
we argue that quantum coherence may provide a mechanism for the preference of configurations which are invariant under 
part of the underlying symmetries.    
A number of models of quantum gravity assume that spaces (spacetimes) are generated from the dynamics of certain  `building-blocks'. 
The number of building-blocks of a macroscopic space is generally huge, and 
their dynamics would thus be complicated and chaotic. 
On the other hand, there exist various symmetries in nature 
such as gauge symmetries, Lorentz invariance, de Sitter symmetry, and so on.
It is a difficult question to see how such symmetries can emerge in general 
from such probably chaotic dynamics of building-blocks. 
This would also be tightly related to the dynamics of the emergence of macroscopic spacetimes, since the de Sitter symmetry should be present almost instantly after the Big Bang (and also approximately in the present era), according to cosmological models \cite{Baumann:2009ds}. 
Of course, it would be natural to impose that theories of quantum gravity should have
underlying symmetries which can interchange 
building-blocks.  
However, it would be difficult to
devise an explicit scenario of spontaneous breaking of the underlying symmetry
to the existing symmetries in nature, 
because there exists a huge hierarchy between them.
In this paper, we consider the problem rather in an opposite direction.
Considering the recent emphasis on Lorentzian treatment in quantum gravity,
we propose that quantum coherence may provide a mechanism to highlight symmetric configurations
over generic non-symmetric ones.

The phenomenon itself is quite common in a broad range of physics. One can find a similar phenomena even in high school text books. 
One example is X-rays diffracted by a crystal. This system has strong peaks in particular directions, each of these peaks represents
particular discrete translational symmetries of the crystal. Though our formalism presented in this paper 
is described in a different manner suited for our purpose,
the aspect is very similar to the text book phenomenon: strong peaks appear in association with 
part of the underlying symmetries.  

The main purpose of this paper is to argue that, due to collective quantum coherence,
peak patterns like the ones mentioned above associated to parts of 
the underlying symmetries 
can actually occur in a large variety of systems, and we explicitly show the phenomenon in a model
tightly related to quantum gravity.
After presenting the general framework of the mechanism in a manner suitable for our purpose,  
we consider the grand partition function of the randomly connected tensor network \cite{Sasakura:2014zwa,Sasakura:2014yoa,
RevModPhys.80.1275} with a Lorentzian modification:
the exponent of an integrand is adjusted to be purely imaginary.
It is known that the grand partition function of the randomly connected tensor network 
has a similar expression as the exact wave functions  \cite{Narain:2014cya} of  
the canonical tensor model \cite{Sasakura:2011sq,Sasakura:2012fb}, 
which is a tensor model \cite{Ambjorn:1990ge,Sasakura:1990fs,Godfrey:1990dt} 
in the Hamilton formalism and studied as a model of quantum gravity. 
Though our final goal would be to use this mechanism in the context of the canonical tensor model, 
we restrict ourselves in this paper to general arguments and concrete demonstrations in simpler related situations 
and leave the application to the canonical tensor model for future study.

This paper is organized as follows. In Section~\ref{sec:mechanism}, we provide the general argument of the mechanism in a 
setup which is common in a broad range of systems with underlying symmetries.
We qualitatively argue that collective quantum coherence highlights configurations invariant under part of underlying 
symmetries of a system, and discuss some conditions which enhance the phenomenon.
In Section~\ref{sec:model}, we explain the model we consider as an example.
It is defined by a Lorentzian modification of the grand partition function of the randomly
connected tensor network, which uses a tensor acting on an $N$-dimensional vectorspace.
In Section~\ref{sec:Neq3}, we consider the $N=3$ case of the model, and find
strong peaks on $SO(2)$ symmetric configurations. 
In Section~\ref{sec:Neq4}, the $N=4$ case is considered. In this case, it turns out that 
there are two possibilities of highlighted symmetries,
$SO(2)$ and $SO(3)$. We observe strong peaks on the configurations with those symmetries similarly to the results for $N=3$. 
We also observe charge quantization in the case of $SO(2)$.
The final section is devoted to a summary and future prospects.
In the appendix, we show some details of the computations in the $N=3$ case.

\section{The mechanism}
\label{sec:mechanism}

The setup of our mechanism is 
 common in physics. 
We consider a quantity which can schematically be represented by
 \[
\Psi(Q)=\int_{\cal C} d\phi\ e^{i S(\phi,Q)},
\label{eq:psi}
\]
where $\phi$ and $Q$ represent sets of multiple variables.
Such quantities appear as path integrals, partition functions, wave functions, etc., 
with corresponding appropriate expressions of $\phi,Q,S(\phi,Q)$, and $\int_{\cal C} d\phi$. 
For instance in the context of quantizing general relativity, a wave function of a universe 
could be expressed in this form by considering $\int_{\cal C} d\phi$ and $Q$ as integration over metrics on manifolds
and metrics on their boundaries respectively. 
Later in this paper, we consider a discrete model 
defined by an expression like \eq{eq:psi}. 
For our purpose, it is essentially important that the integrand of \eq{eq:psi} takes complex values
with non-trivial phases in general.  
Thus, our setup must have Lorentzian characteristics rather than Euclidean:
the integrand of \eq{eq:psi} should not take just positive real values.

The class of theories we consider are assumed to have group symmetries in the following sense.
For convenience, let us discretely label the variables $\phi,Q$ as $\phi_a\ (a=1,2,\cdots,N)$ and $\ Q_i\ (i=1,2,\cdots,M)$ 
respectively. In general it is also possible to consider continuous labels such as spacetime coordinates. 
Let us assume that group transformations of $\phi$ and $Q$ are given by
\[
\begin{split}
(\phi_g)_a&=R(g)_{a}{}^b\phi_b,\\
(Q_g)_i&=\tilde R(g)_i{}^{j} Q_j,
\end{split}
\]
where $R(g)$ and $\tilde R(g)$ are representations of a group $G$ for elements $g$. 
For our purpose, it is necessary that neither of the %
representations, $R$ nor $\tilde R$, are trivial.
We do not assume $R$ or $\tilde R$ to be irreducible, and they are 
allowed to contain trivial representations in general. 
We assume that the integration and $S$ are invariant as 
\[
\begin{split}
\int_{\cal C} d \phi_g&=\int_{\cal C} d\phi, \\
S(\phi_g,Q_g)&=S(\phi,Q),
\end{split}
\label{eq:invariance}
\]
for any $g\in G$.
In particular, the assumptions imply the invariance of $\Psi$ as
\[
\Psi(Q_g)=\Psi(Q).
\]

Now, 
an interesting physical question would be what the locations of maxima of $|\Psi(Q)|^2$ are.
For the simplicity of the following discussions, let us assume $S(\phi,Q)$ to be real. This 
is common in physics, but the assumption may not be necessary 
as long as the integrand can take various complex values with non-trivial phases.
To illustrate the mechanism, let us consider a crude approximation of \eq{eq:psi}. This approximation is based on the principle of stationary phase. The principle states that the major contribution of a highly oscillatory integral comes from points where the phase function (in our case $S(\phi, Q)$) is stationary \cite{Wong:2001}.
This approximation would thus be given by only picking up stationary points as 
\[
\Psi(Q)\sim \sum_{\sigma=1}^{n_{crit}} A_\sigma e^{i S(\phi^\sigma,Q)},
\label{eq:critical}
\]
where $\phi=\phi^\sigma\ (\sigma=1,2,\cdots,n_{crit})$ are the critical points defined by
\[
\left. \frac{\partial S(\phi,Q)}{\partial \phi_a} \right |_{\phi=\phi^\sigma}=0,\ \ (a=1,2,\cdots,N),
\label{eq:criticalpoints}
\]
and $A_\sigma$ are the prefactors obtained by performing Gaussian integrations around each critical 
point\footnote{The expression with the sum over critical points can be sophisticated  
by a method based on the Picard-Lefschetz theory. For instance, 
see Section 3 of \cite{Witten:2010cx} and references therein for more details.}.
Here, for simplicity, it is assumed that the critical points are isolated for generic values of $Q$. 
If $M$ and $N$ are 
sufficiently large and $S(\phi,Q)$ is not too 
simple\footnote{For instance, $S(\phi,Q)$ should not be a quadratic function of $\phi$ and $Q$.}, 
 expression \eq{eq:critical} suggests that $\Psi(Q)$ has the following properties.
One is that, for generic values of $Q$, there would exist 
 a substantial number, $n_{crit}$, of critical points, and 
there would be no strong correlations among values of phase $S(\phi^\sigma,Q)$ of different critical points.
Therefore, $\Psi(Q)$ will be suppressed by mutual cancellations among the contributions of all the critical points.
It is also likely that, for generic values of $Q$, $\Psi(Q)$ behaves almost randomly as a function of $Q$ due to 
the uncorrelated values of $S(\phi^\sigma,Q)$ among the critical points. 
These properties would make it unlikely that there exist sensible observables 
in such generic regions of $Q$.
Though such situations would not be generally true for all the possible cases,
the kinds of cancellations and randomness mentioned above  
are expected to naturally occur, when $M$ and $N$ (and $n_{crit}$ as well) are sufficiently large 
and $S(\phi,Q)$ is not too simple.\footnote{This aspect seems to have a certain philosophical similarity 
with the random dynamics developed in \cite{Bennett:1986xb}, since any sophisticated discrete model would give some preferred symmetric states. There is a difference, however, since in our mechanism the model  which is chosen 
will influence the symmetries 
which are highlighted, whereas random dynamics dictates that the specific model
should not matter much for the emergent symmetries at large scales.
It would be interesting to compare our mechanism and theirs to a deeper extent.} 

On the other hand, there exist values of $Q$ where $|\Psi(Q)|^2$ seems to take larger values. 
Let us consider a value of $Q$, say $Q^H$, which is invariant under a subgroup $H$ of $G$:
\[
Q^H_{h}=Q^H,\ {}^\forall h \in H \subset G.
\label{eq:invP}
\]
Then, for such a $Q^H$, critical points exist along the trajectories of the group action $H$:
\[
(\phi^{\sigma}_h)_a&=R(h)_{a}{}^b(\phi^{\sigma})_b,\ {}^\forall h \in H \subset G.
\]
This is an immediate consequence of the invariance in \eq{eq:invariance}.
The representation $R(H)$ is generally reducible, and therefore a critical point $\phi^{\sigma}$ may be contained in the 
trivial part of the representation. 
 In that case, the critical point is usually an isolated single point 
(except if it happens to coincide with another critical point, which might or might not have some deeper reasons).  
Otherwise, they form a set of critical points, on which the group action $H$ is non-trivially represented. 
For the 
simplicity of our discussions below, let us assume both $H$ and $G$ to be Lie groups. The discussions below can obviously be generalized to finite groups and a similar mechanism will hold if
the orders of the groups are large enough. 
In the case of Lie groups, the set of critical points on which $H$ acts non-trivially form a continuous set of critical points. 
Taking into account the differences of the two classes of critical points, $\Psi(Q^H)$ can now be represented by
\[
\begin{split}
\Psi(Q^H)\sim \sum_{\sigma: \hbox{\tiny non-trivial}} \int_H dh \left| \frac{\partial \phi}{\partial h} \right| A_\sigma e^{iS(\phi^\sigma ,Q^H)}\\+
\sum_{\sigma:\hbox{\tiny trivial}}A_\sigma e^{iS(\phi^\sigma ,Q^H)},
\label{eq:contpsi}
\end{split}
\]
where the first term denotes the contributions from the continuous critical points, while the latter contains the contributions from the 
isolated ones. Here $A_\sigma$ in the first term is determined by the Gaussian integrations over the transverse 
directions to the group trajectories. 
One important matter is that the phase function is constant along each group trajectory 
because
$S(\phi^\sigma_h ,Q^H)=S(\phi^\sigma_h ,Q^H_h)=S(\phi^\sigma ,Q^H)$ for $^\forall h\in H$.
The constancy of the phase similarly holds 
even when the prefactor $A_\sigma$ is included.
This is actually an exact property beyond the Gaussian approximation above,
because the whole system of $\phi$ is invariant under $H$ for $Q=Q^H$.

In \eq{eq:contpsi}, there exist two main reasons for which $|\Psi(Q^H)|^2$ may become relatively large due to the 
continuous critical points. 
One is that phases are constant along each group trajectory. Therefore, continuous 
critical points on each trajectory contribute coherently to $\Psi(Q^H)$.
The other is that, when $Q$ reaches $Q^H$, some of the isolated critical points are connected by 
a group orbit generated by $H$ to form a continuous set of critical points. 
Therefore, the number of critical points is enhanced from a finite number to a continuous infinite. 
No divergences are generally caused by that though for compact groups, as can be seen in the expression \eq{eq:contpsi}.

Expression \eq{eq:contpsi} suggests some conditions for large enhancement.
One is that the trajectories of continuous critical points should occupy large subspaces in the space of $\phi$.
Another is that the number of
sets of continuous critical points should not be too large to avoid mutual cancellations among them. 
These conditions may be rephrased in that there exist a small number (ideally one) of continuous sets of 
critical points 
whose orbits $R(H)\phi^\sigma$ have large volumes.
As for the dimension of the subgroup $H$, there seems to exist a tension between 
the following two factors, and it seems hard to say which factor wins.
One is that configurations invariant under smaller dimensional subgroups are more common,
because it is easier for configurations to satisfy the symmetry condition \eq{eq:invP} for smaller
dimensional subgroups.
On the other hand, enhancement tends to become larger for larger dimensional subgroups, 
because the dimensions of group orbits in the space of $\phi$ are larger. 

The preferences discussed above would be consistent with the present form of the actual universe. 
Comparing to the large supposed symmetry which can interchange `building-blocks' of spaces, 
the existing symmetries in nature are extremely small.
On the other hand, the representation space of the existing symmetries is the universe itself (for instance, consider 
the translational symmetry), and therefore the dimension of the representations is 
extremely large: it should be
in the order of the number of `building-blocks' of the universe.

Lastly, we want to comment on an essential difference between the Lorentzian treatment and the Euclidean one. 
In the Euclidean treatment, preferred configurations are usually obtained by minimization of $S(\phi,Q)$,
often referred to as ground states.
Here, whether preferred configurations are symmetric or not is just an outcome, but is not relevant in 
determining them.
On the other hand, in the Lorentzian treatment, 
quantum coherence is the essence rather than values of $S(\phi,Q)$, 
and symmetries play important roles in determining preferred configurations. 
This seems to imply an important paradigm change from Euclidean to Lorentzian: 
The collective behavior of the critical points is more important than the value of a single special point.

\section{The model}
\label{sec:model}
Our final aim for developing this formalism 
is to apply the highlighting mechanism discussed in the previous section
to the canonical tensor model \cite{Sasakura:2011sq,Sasakura:2012fb}, a model of quantum gravity, 
to explore the possibility of emergence of macroscopic spacetimes.
On the other hand, the mechanism itself will generally hold in various settings with underlying symmetries.
Therefore, to concentrate on the mechanism itself,
we restrict ourselves to show the validity of the mechanism in some concrete simple examples connected to the canonical
tensor model: we consider 
the grand partition function of the randomly connected tensor networks \cite{Sasakura:2014zwa,Sasakura:2014yoa,
RevModPhys.80.1275} with a Lorentzian modification. 
In fact, this setting is not so far from the canonical tensor model \cite{Sasakura:2011sq,Sasakura:2012fb}, 
since its known exact wave functions \cite{Narain:2014cya} have similar expressions.

The grand partition function \cite{Sasakura:2014zwa} of the randomly connected tensor network is defined by 
\[
Z_{T}(Q)=\int_{\cal C} d\phi\ e^{S_T(\phi,Q)},
\label{eq:grand}
\]
where the integration region ${\cal C}$ is tilted from a real plane to a complex one
to make the integration convergent \cite{Sasakura:2014zwa},
and 
\[
&d\phi=\prod_{a=1}^N d\phi_a,\\
&S_T(\phi,Q)=\phi^2+Q\phi^3\label{eq:defofS}
\]
with short-hand notations, $\phi^2=\phi_a\phi_a$ and $Q\phi^3=Q_{abc}\phi_a \phi_b \phi_c$. 
Here $Q_{abc}$ is assumed to be real and totally symmetric.

In this paper, we rather discuss a quantity which is similar but has a Lorentzian modification
of multiplying the exponent by $i$, 
\[
\Psi_{T}(Q)=\int_{\mathbb{R}^N} d\phi\ e^{i S_T(\phi,Q)-\epsilon \phi^2},
\label{eq:defofpsip}
\]
where $\mathbb{R}^N$ denotes the whole $N$-dimensional real space, 
and $\epsilon>0$ is a regularization parameter assuring the convergence of the integration for
any real values of $Q$. In addition to the interest in the exact wave functions of the canonical tensor model, 
$\Psi_{T}(Q)$ may be regarded as 
the grand partition function of the randomly connected tensor network with 
imaginary parameters.

Note that $\Psi_{T}(Q)$ is obviously symmetric under the orthogonal group $G=O(N)$, 
and satisfies the conditions required in the previous section, where the representations are given by
\[
\begin{split}
(\phi_g)_a&=R(g)_{a}{}^b \phi_b, \\
(Q_g)_{abc}&=R(g)_{a}{}^dR(g)_{b}{}^eR(g)_{c}{}^f Q_{def}
\end{split}
\]
with $R(g)_a{}^b$ being the matrix of the ($N$-dimensional) vector representation of $G=O(N)$ for $g\in G$.
The regularization parameter $\epsilon$ will generally be taken to be small positive values, 
but we will not discuss the details of its vanishing limit $\epsilon \rightarrow 0^+$.
This is because $\Psi_{T}(Q)$ may have some singular behaviors in the limit $\epsilon \rightarrow 0^+$,
when some of the components of $Q$ vanish.
Analyzing the details of such behavior 
is not in the range of our main purpose of this paper: 
we are rather interested in
showing the enhancement of \eq{eq:defofpsip} for symmetric $Q$, and, to see this, it is enough to 
take $\epsilon$ sufficiently small. 

\section{Simplest non-trivial case: $N=3$}
\label{sec:Neq3}
 We will now discuss a concrete example of the mechanism at work. We will use the model defined by \eqref{eq:defofpsip} in the case of $N=3$. In this case \eqref{eq:defofpsip} is invariant under transformations of the orthogonal group $G=O(3)$, and the totally symmetric real tensor $Q_{abc}$ has 10 independent parameters. As mentioned above \eqref{eq:invP} we will now consider a subgroup $H\subset G$. The subgroup $H=SO(2)$ will be used with the following representation of its Lie algebra:
 \begin{equation}
    T = \begin{pmatrix} 0 & 1 & 0 \\ -1 & 0 & 0 \\ 0 & 0 & 0\end{pmatrix}.\label{eq:N=3:T_matrix}
 \end{equation}
The generator of $H=SO(2)$ can always be put into this form by performing an appropriate $G=O(3)$ transformation. 
 Demanding \eqref{eq:invP} to hold leads to the following requirement for $Q$:
 \begin{equation}
     T_a{}^{a'} Q_{a'bc} + T_b{}^{b'} Q_{ab'c} + T_c{}^{c'} Q_{abc'} = 0.\label{eq:N=3:T_req}
 \end{equation}
This restricts the tensor a lot, as there are only two independent parameters left:
 \begin{align}
      &Q_{113} = Q_{223} \equiv \frac{x}{3},\nonumber\\
      &Q_{333} \equiv y,\label{eq:symm:toy:N=3:symm}\\
      &\text{others} = 0.\nonumber
  \end{align}
  \eqref{eq:defofS} is now written as
  \begin{equation}
      S(\phi, x, y) =  (\phi_1^2 + \phi_2^2) (1+ x \phi_3) + \phi_3^2 + y \phi_3^3.\label{eq:N=3:S_symmetric}
  \end{equation}
  The critical points can now easily be computed using \eqref{eq:criticalpoints}. There are two isolated critical points 
  and a continuous set of critical points which is invariant under the group action of $H$:
  \begin{align}
      &\phi_1= \phi_2 = \phi_3 = 0\label{eq:N=3:crit_point_1},\\
      &\phi_1 = \phi_2 = 0;  \phi_3 = -\frac{2}{3y}\label{eq:N=3:crit_point_2} ,\\
      &{\phi_1}^2 + {\phi_2}^2 = \frac{2x-3y}{x^3}\equiv r_0^2 ;\ \phi_3 = - \frac{1}{x}\label{eq:N=3:crit_point_S1}.
  \end{align}
  So for $x>0, 2x > 3y $ or $ x<0, 2x<3y$ there is a continuous set of critical points with topology 
  $S^1$.\footnote{Complex valued critical points may also be relevant for the evaluation of $\Psi$
  according to the aforementioned method based on Picard-Lefschetz theory. 
  Hence, in principle, the continuous critical points may become relevant even in the outside of these regions 
  bounded by the inequalities. 
  However, such contributions do not seem important in our examples, as we will see explicitly.} 
  If we make $r_0$ large by taking $x\sim 0$ within these regions with finite $y$,
  the continuous set of critical points will occupy a relatively large subspace in the space of $\phi$,
  and $|\Psi(Q)|^2$ will become large.
 
To see this explicitly, we can compute the contributions of the critical points by performing the Gaussian integrations 
around them, corresponding to \eq{eq:contpsi}.  
For large $r_0$ we obtain
  \begin{align}
   \begin{split}\Psi(x,y) &\approx \pi^{3/2} e^{i 3 \pi/4} \left(1 + i\left(\frac{3y}{2x - 3y}\right) e^{i \frac{4}{27 y^2}}\right)\\
 			  &\ \ \ \ +\frac{2\pi^2}{|x|} e^{i \frac{x-y}{x^3}}\delta_{\text{sgn}(x), -\text{sgn}(y)},
   \label{eq:sumofcritical}
   \end{split}
  \end{align}
  where $\text{sgn}(x)$ denotes the sign of $x$.
  The explicit calculation of this can be found in the appendix. The last term is the contribution
  of the continuous set of critical points \eq{eq:N=3:crit_point_S1}, and clearly dominates over the isolated critical points for large $r_0$ (or $x\sim 0$). 
  The second term appears to have a pole at $2x-3y=0$, 
but this is just a consequence of the approximate treatment with the Gaussian integration. Indeed there is
no such a singular behavior of $\Psi$, as can be checked by an exact formula below.
  
  Because of the simplicity of the current model, it is also possible to obtain an exact expression of the 
  important contribution. The calculation is shown in the appendix, and the result is
  \begin{equation}
      \Psi(x,y) = i \pi \int_{\mathbb{R}} d\phi_3 f(x,y,\phi_3+i\lambda) + \frac{2\pi^2}{\vert x\vert} e^{i\frac{x-y}{x^3}}\delta_{\text{sgn}(x), -\text{sgn}(y)}.
      \label{eq:exactform}
  \end{equation}
  Here $f(x,y,\phi_3)$ is defined in \eq{eq:N=3:Psiintegral},
  and $\lambda$ is an arbitrary real number with $\text{sgn}(\lambda)=\text{sgn}(y)$.
 This is obtained by considering a contour integration shown in Fig.\ref{fig:N=3:contour}
 to single out the contribution of a pole. This expression does not anymore depend on the regularization 
 parameter $\epsilon$, since the first term is a non-singular convergent integration 
 with an exponential damping behavior of the integrand $\sim \exp(-3 y \lambda \phi_3^2)$. 
The second term is exactly the contribution of the continuous critical points for large $r_0$, so the stationary phase approximation 
above seems to work really well for a large space of critical points.
  
  \subsection{Asymmetric configuration}
   We showed that for small $x$ the behavior of $\Psi(x,y)$ is dominated by the continuous critical points. We will now show that the symmetric configuration is the preferred configuration. 
   
   Let us first note that the situation of \eqref{eq:N=3:crit_point_S1} is rather special. If $Q$ does not have a certain symmetry, then such a continuous set of critical points will in general not exist. In the case of $N=3$ it can actually be shown explicitly that 
 any other additional non-symmetric 
   terms in \eqref{eq:N=3:S_symmetric} will remove the continuous set of critical points in favor of a number of isolated points. If an asymmetric term is parametrized by $z$, 
   taking the limit $z\rightarrow 0$ moves these isolated points towards the circle defined by \eqref{eq:N=3:crit_point_S1},
   and for $z=0$ they are linked by a constant trajectory of $S$, which forms the circle of continuous critical points.
   
   From this and earlier discussions we expect that adding an asymmetric term disrupts the symmetry and 
   $\Psi$ will be suppressed. The simplest way to disrupt the symmetry would be to allow $Q_{113}$ and $Q_{223}$ to take different values:
   \begin{align}
    &   Q_{113} \equiv \frac{x_1}{3} \neq Q_{223} \equiv \frac{x_2}{3},\nonumber\\
     &  Q_{333} \equiv y,\label{eq:N=3:asymm:ex1}\\
     & \text{others} = 0.\nonumber
  \end{align}
  The integral is then given by
  \begin{equation}
      \Psi(x_1,x_2,y) = \int_{\mathbb{R}^3} d\phi \ e^{i (\phi_1^2 + \phi_2^2 + \phi_3^2 + x_1 \phi_1^2 \phi_3 + x_2 \phi_2^2 \phi_3 + y \phi_3^3) - \epsilon \phi^2}.
  \end{equation}
  After performing the Gaussian integration over $\phi_1$ and $\phi_2$ one finds
  \begin{align}
      \Psi(x, z , y) = i \pi \int_{\mathbb{R}} d\phi_3 \, \frac{e^{i (\phi_3^2+y \phi_3^3) - \epsilon \phi_3^2}}{\sqrt{1+(x+z)\phi_3 + i \epsilon}\sqrt{1+(x-z)\phi_3+ i \epsilon} }.\label{eq:N=3:asymm:psi_asymm_sol_int}
  \end{align}
  Here we have introduced $x = \frac{x_1+x_2}{2}$ and $z=\frac{x_1-x_2}{2}$. This param\-etrization is mainly chosen because one can now identify the symmetric configuration by simply setting $z=0$. For $z=0$ this integral reduces to \eqref{eq:N=3:Psiintegral}. Although it proves quite difficult to solve this integral analytically, 
it can be evaluated numerically quite well. 
   \begin{figure}
   \centering
     \includegraphics[scale=0.35
     ]{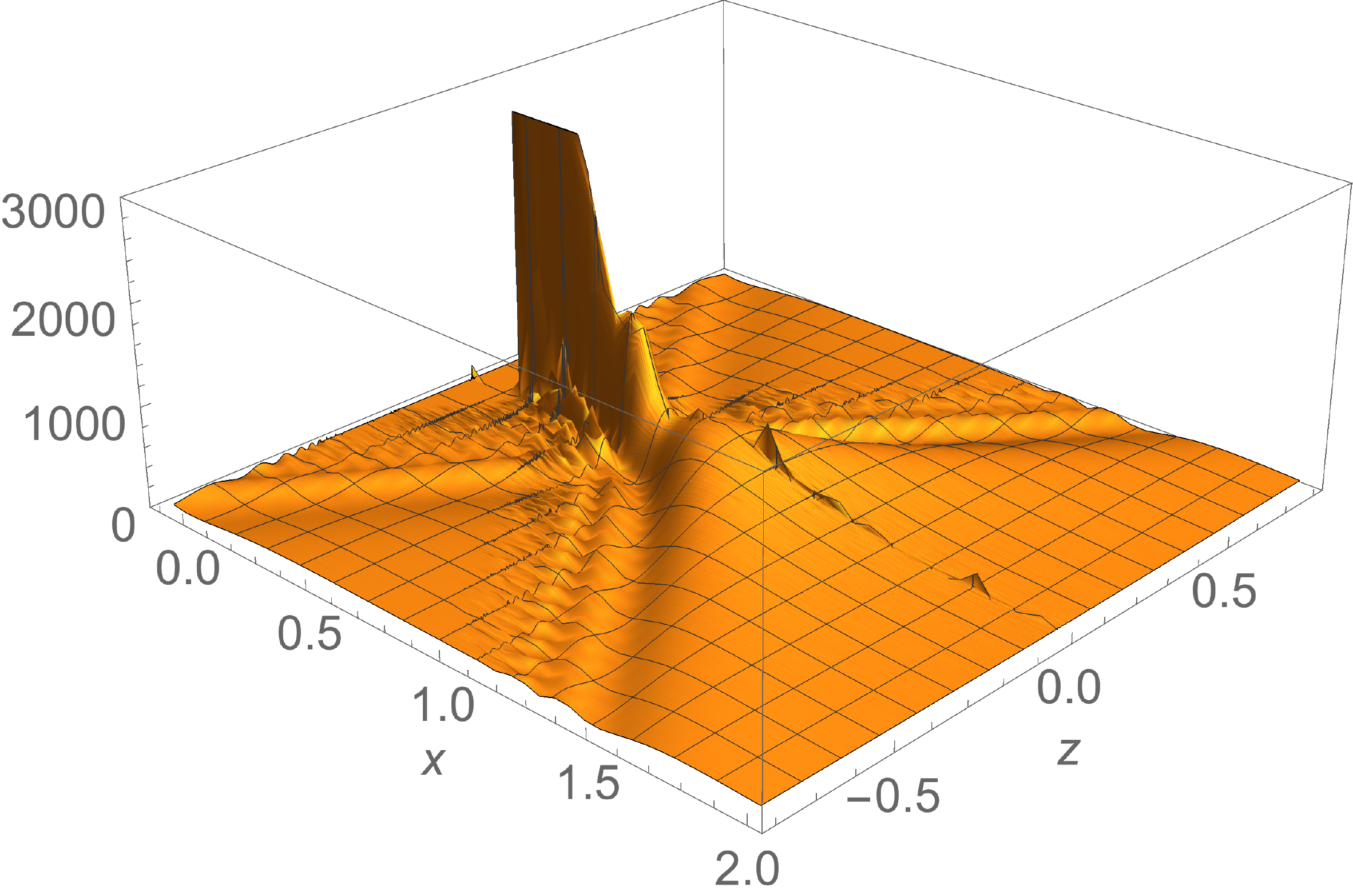}
    \caption{Numerical integration of \eqref{eq:N=3:asymm:psi_asymm_sol_int} 
    with $y=-1, \epsilon=0.0001$.
The figure plots the values of $|\Psi(x,z)|^2$ against $x,z$. 
Values over 3000 are chopped.
For $z$ we used a step of $\Delta z = 0.001$ and for $x$ a step of $\Delta x = 0.05$.
The function peaks around $z=0$, which agrees with the expectation from the highlighting mechanism. For $x\rightarrow 0^+$ the function diverges as expected from \eq{eq:sumofcritical} or \eq{eq:exactform}. 
}
    \label{fig:N=3:Q113neqQ223}
  \end{figure}
  
  As can be seen in Fig.\ref{fig:N=3:Q113neqQ223}, $|\Psi|^2$ clearly peaks at $z=0$, which indicates that the system prefers  symmetric configurations.
We also studied the region $x<0$ with $y=-1$ numerically, but have not observed any interesting behavior.
These seem to confirm our expectations, as there are either no real solutions to \eqref{eq:N=3:crit_point_S1} in that case, or the solution has very small $r_0$.   
  
The above numerical study is just one of the possible asymmetric terms, one more will now be investigated before making some more general (but local) statements. Instead of \eqref{eq:N=3:asymm:ex1} we now take
  \begin{align}
      Q_{113} &= Q_{223} \equiv \frac{x}{3},\nonumber\\
      Q_{333} &\equiv y,\label{eq:N=3:Q122}\\
      Q_{122} &\equiv \frac{z}{3}.\nonumber
  \end{align}
  All other terms (of course with the exception of permutation of indices) are again put to zero. The following integral now needs to be solved: 
  \begin{equation}
      \Psi(x,y,z) = \int_{\mathbb{R}^3} d\phi 
      \ e^{i(\phi_1^2(1+x \phi_3)+\phi_2^2(1+x \phi_3 + z\phi_1) + \phi_3^2 + y \phi_3^3) - \epsilon \phi^2}.
  \end{equation}
  The $\phi_2$ integration can be done readily, which gives
  \begin{equation}
      \Psi(x,y,z) = e^{i \pi/4} \sqrt{\pi} \int_{\mathbb{R}^2} d\phi_1 d\phi_3 \frac{e^{i (\phi_1^2(1+x \phi_3) + \phi_3^2 + y \phi_3^3)- \epsilon(\phi_1^2+\phi_3^2)}}{\sqrt{1+x \phi_3 + z\phi_1 + i \epsilon}}.\label{eq:N=3:PsiQ112}
  \end{equation}
  \begin{figure}
   \centering
     \centering
     \includegraphics[scale=0.3
     ]{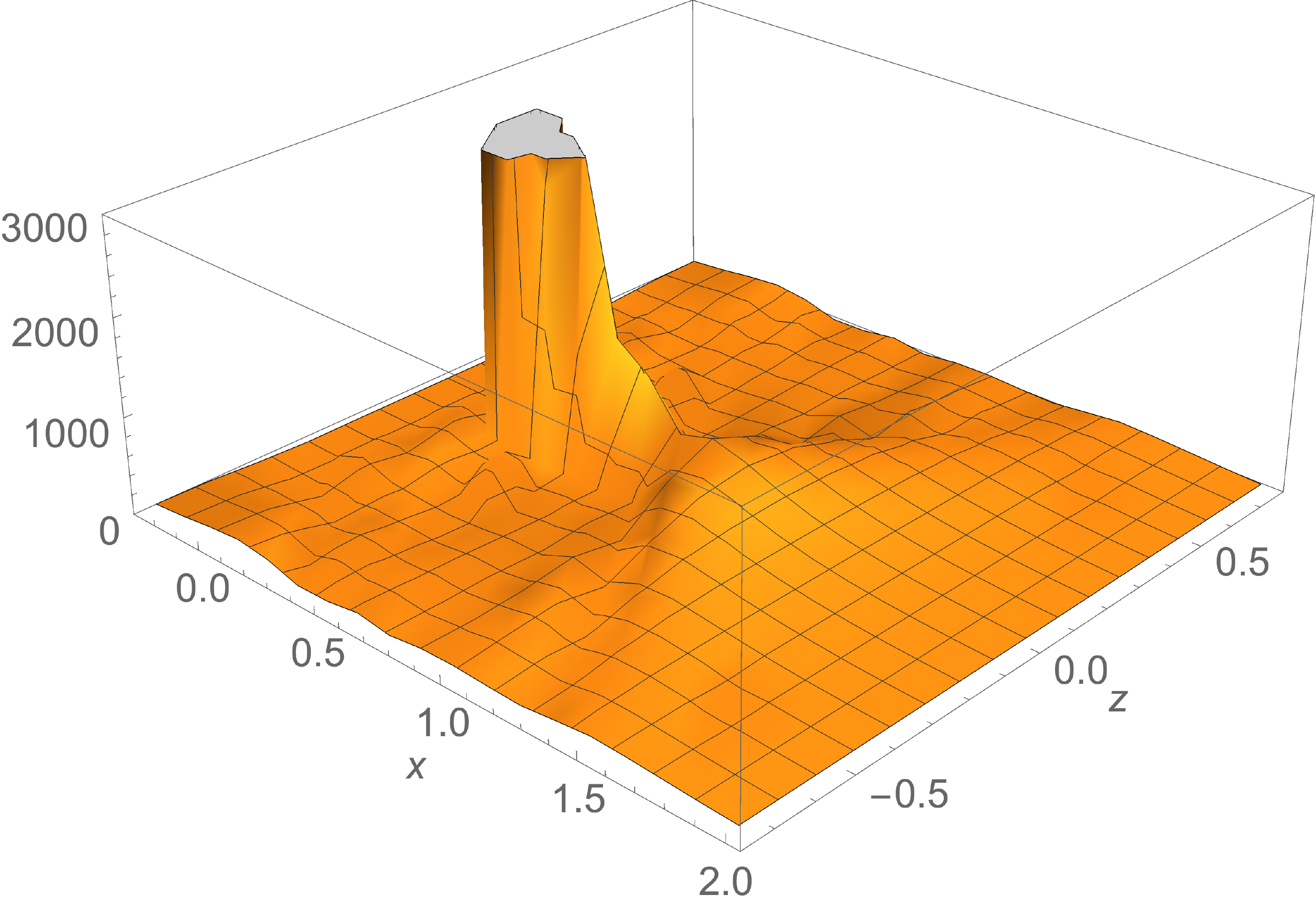}
    \caption{Numerical integration of the integral in \eqref{eq:N=3:PsiQ112} with $y=-1, \epsilon=0.0001$.
The figure plots the values of $|\Psi(x,z)|^2$ against $x,z$. Values over 3000 are chopped. 
For $z$ we used a step of $\Delta z = 0.125$ and for $x$ a step of $\Delta x = 0.125$.
The function peaks around $z=0$, which agrees with the expectation from the highlighting mechanism. For $x\rightarrow 0^+$ the function diverges as expected from \eq{eq:sumofcritical} or \eq{eq:exactform}. }
    \label{fig:N=3:Q112}
  \end{figure}%
   This double integration can be done numerically and the result is shown in Fig.\ref{fig:N=3:Q112}. The result is similar to 
  Fig.\ref{fig:N=3:Q113neqQ223}, 
  supporting the generality of the phenomenon.
  
 \subsection{Perturbations}
By adding certain asymmetric terms, we have seen that the symmetric configurations seem to be preferred. 
To be more general one can look at general small perturbations around the symmetric 
configurations to make local statements. Let the tensor given in \eqref{eq:symm:toy:N=3:symm} be denoted by $Q^0$ and its corresponding action, \eqref{eq:N=3:S_symmetric}, by $S_0$. 
$\Psi(Q)$ can be written, at least formally, for small perturbations $Q=Q^0 + \delta Q$ as
  \begin{equation}
      \Psi(Q^0 + \delta Q) = \Psi(Q^0) + \sum_{n>0} \int_{\mathbb{R}^3} d\phi \frac{1}{n!}(i\delta Q \phi^3)^n e^{i S_0(\phi)
      -\epsilon \phi^2}. 
  \end{equation}
  Considering the first order term in $\delta Q$ we get an integral like
  \begin{equation}
      \int_{\mathbb{R}^3} d\phi\, \phi_a \phi_b\phi_c \, e^{i (\phi^2 + x (\phi_1^2+\phi_2^2)\phi_3 + y\phi_3^3)-\epsilon \phi^2}.
  \end{equation}
  Note that any term which is odd in $\phi_1$ or $\phi_2$ will vanish, since the whole integrand is then odd in $\phi_1$ or $\phi_2$. There is actually just one term (which is not present in the unperturbed function) which is not odd but also vanishes:
  \begin{align*}
      &\int_{\mathbb{R}^3}  
      d\phi\, (\phi_1^2 \phi_3 - \phi_2^2 \phi_3) e^{i (\phi^2 + x (\phi_1^2+\phi_2^2)\phi_3 + y\phi_3^3)-\epsilon \phi^2} \\
      &= \int_{\mathbb{R}^3} d\phi \, \phi_1^2 \phi_3 e^{iS_0-\epsilon \phi^2} 
      - \int_{\mathbb{R}^3} d\phi \,  \phi_2^2 \phi_3 e^{iS_0-\epsilon \phi^2}\\
      &=0.
  \end{align*}
  This means that $\left.\partial \Psi/\partial Q_{abc}\right\vert_{Q=Q^0}\delta Q_{abc} = 0$ 
  for any perturbations
  $\delta Q$ transverse to the symmetric configurations.
  Therefore, $\Psi(Q^0)$ is a local extremum with respect to the transverse directions. 
  In general the second derivative will not be negative, as is what one would expect for a local maximum. 
  However, in the region where the continuous critical points are important, $Q=Q^0$ becomes a local maximum of 
  $|\Psi(Q)|^2$ with respect to the transverse directions. 
  This can be qualitatively shown by assuming that the contributions of the continuous set of 
critical points $\phi^\sigma_h$ dominate over the other isolated contributions in the summation \eq{eq:contpsi}:
  \begin{align}
     &\Psi(Q^0)^*\left.\frac{\partial^2 \Psi(Q)}{\partial Q_{abc}\partial Q_{def}}\right\vert_{Q=Q^0} 
     \delta Q_{abc}\delta Q_{def} \nonumber\\
     &= -\Psi(Q^0)^* 
     \int_{\mathbb{R}^N} d\phi\, \left( \delta Q \phi^3\right)^2\, e^{i S_0(\phi, x,y)-\epsilon \phi^2}\nonumber \\
        &\approx -\left\langle \left(\delta Q (\phi^\sigma_h)^3\right)^2 
        \right\rangle_H \vert\Psi(Q^0)\vert^2 < 0,
        \label{eq:delS2}
\end{align}
where we used the short-hand notation $\delta Q\phi^3\equiv \delta Q_{abc} \phi_a \phi_b\phi_c$, and  
$\langle \cdots \rangle_H$ denotes taking an average over the group trajectory of $H$.
With $\left.\partial \Psi/\partial Q_{abc}\right\vert_{Q=Q^0}\delta Q_{abc} = 0$, 
\eq{eq:delS2} proves 
$|\Psi(Q)|^2$ takes a local maximum at $Q=Q_0$ with respect to the transverse directions to the symmetric configurations.
This qualitative proof of the relation between the classicality of $\phi$ and the maxima
of $|\Psi|^2$ is generally valid 
for any symmetric configurations and $N$.   
Here, the existence of the imaginary unit in the exponent is essentially important
for the appearance of the minus sign in \eq{eq:delS2}, and therefore the relation should be considered to be 
a characteristic of the Lorentzian treatment.
The stationary points of the action are often regarded as the `classical path' of a system, and 
the above argument shows that 
in the `classical limit' the system generally prefers symmetric configurations. 
  
\section{$N=4$}
\label{sec:Neq4}
 In the case of $N=3$ there was only one possible Lie subgroup which could function as a partial symmetry. In the case of $N=4$, 
 there are three possibilities: $SO(2)$, $SO(2)\times SO(2)$ and $SO(3)$. However, it will turn out that the $SO(2) \times SO(2)$ reduces to be either trivial or an $SO(2)$. 
In what follows we will first consider $SO(2) \times SO(2)$ and $SO(2)$.
 
By $O(N)$ rotations, the $SO(2) \times SO(2)$ generators for $N=4$ can be put in the following representation:
 \begin{equation}
  T = \begin{pmatrix} 0 & t_1 & 0 & 0 \\ -t_1 & 0 & 0 & 0 \\ 0 & 0 & 0 & t_2 \\ 0 & 0 & -t_2 & 0 \end{pmatrix}.\label{eq:N=4:T_SO(2)}
 \end{equation}
 Here $t_1$ and $t_2$ are parameters with $t_1>0$ and $t_2\geq0$, 
 so \eqref{eq:N=4:T_SO(2)} represents two independent generators in general. 
 However, it turns out that in order to solve \eqref{eq:N=3:T_req} there are only two possible values of $t_2$: $t_2=0$ or $t_2=2t_1$\footnote{There is also a solution for $2t_2 = t_1$, but this leads to no new results.}. Other values of $t_2$ lead
to the trivial configuration ($Q_{abc}=0$). 
In each case, $t_2=0$ and $t_2=2t_1$, \eq{eq:N=4:T_SO(2)} represents only one $SO(2)$ generator,
and therefore the symmetry is $SO(2)$, not $SO(2) \times SO(2)$.
This preference of `charge quantization' is generally true for any $N$: 
one can find more solutions of symmetric configurations of $Q$ 
by such `charge quantization'  restricting the freedom of Lie group generators.
  
 For $t_2=0$ the following symmetric tensor solves \eqref{eq:N=3:T_req}:
 \begin{align}
 \begin{split}
&  Q_{113} = Q_{223}=x_1, Q_{114}=Q_{224}=x_2,\\
&  Q_{333}=y_1, Q_{444}=y_2, Q_{334}=z_1, Q_{344}=z_2.
 \end{split}
 \end{align}
 All other terms (of course with the exception of permutation of indices) are again put to zero. The same methods as before can now be used to find the critical points. The specific expressions are nontrivial and not very relevant for the discussion, but the critical points again admit a solution of the form $\phi_1^2 + \phi_2^2 = r_0^2, \phi_3=\text{constant}, \phi_4=\text{constant}.$ This is also what we expect, since the critical points need to admit an SO(2) symmetry
 represented as the rotations of $\phi_{1,2}$. 
 
 For $t_2=2t_1$ there is a more restricted set of parameters:
 \begin{align}
 \begin{split}
  Q_{114}=-Q_{224} &= - Q_{123},\\
  Q_{113}=-Q_{223} &= Q_{124}.
 \end{split}
 \end{align}
 All other terms (of course with the exception of permutation of indices) are again put to zero. This gives, besides 
 some isolated critical points, also a set of continuous critical points 
 satisfying $\phi_1^2 + \phi_2^2 = 2(\phi_3^2 + \phi_4^2) = r_0^2$. 
 
We have performed some numerical investigations under some perturbations around the symmetric configurations 
for the above cases, too. We have obtained similar results as the $N=3$ case: strong peaks on the symmetric configurations.   
 
\begin{figure}
   \centering
     \includegraphics[scale=0.5
     ]{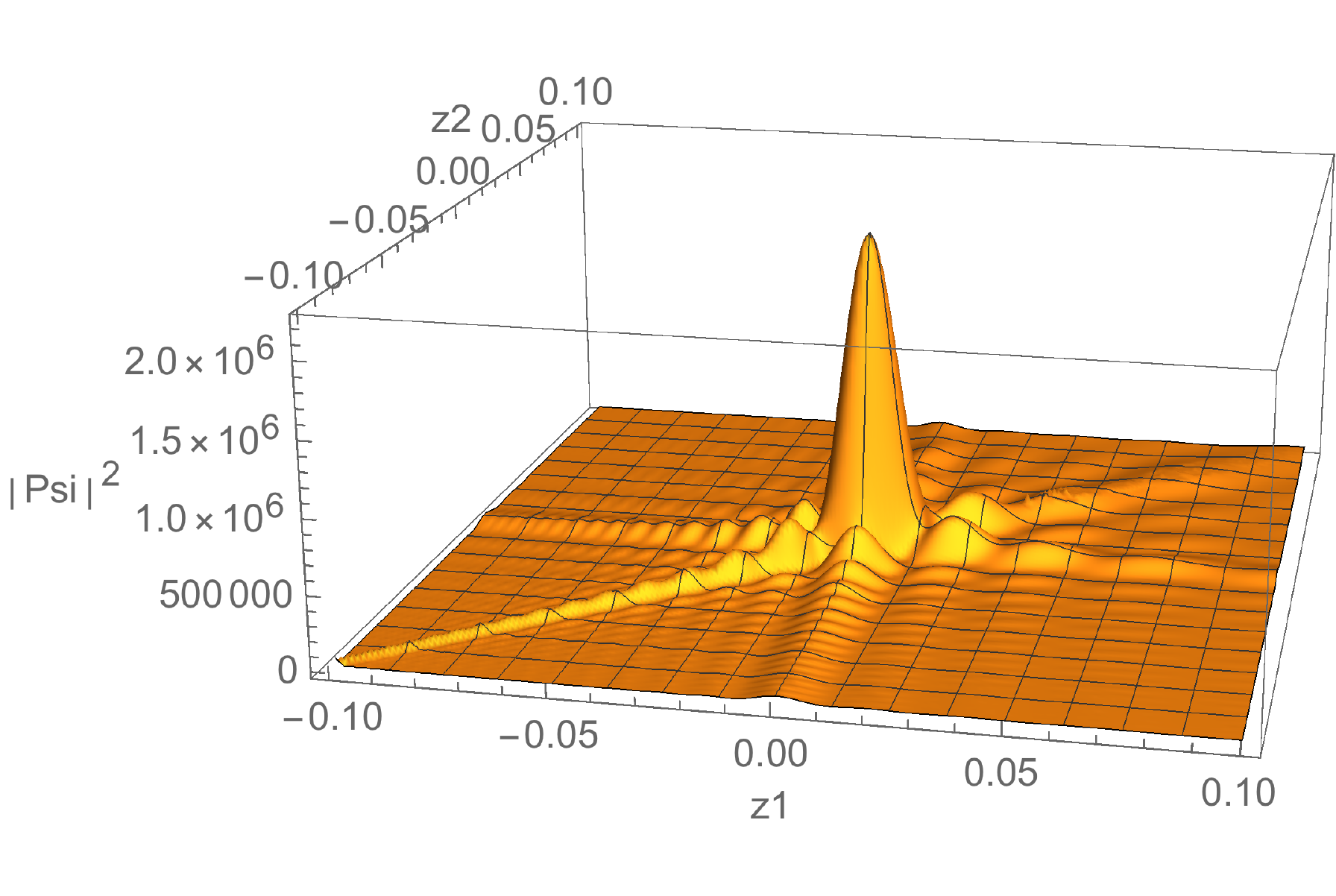}
    \caption{Numerical integration of the integral in \eq{eq:Neq4SO3} with $x_1=0.3, y=-1, \epsilon=0.0001$.
The figure plots the values of $|\Psi(z_1,z_2)|^2$ against $z_1,z_2$, where $x_2=x_1+z_1,x_3=x_1+z_2$. 
We used a step of $\Delta z_{1,2} = 0.001$.
The function peaks around $z_1=z_2=0$, which is the $SO(3)$ peak, and there are also smaller $SO(2)$ peaks
along $z_1=0$, $z_2=0$, and $z_1=z_2$.
}
    \label{fig:N=4SO(3)}
  \end{figure}

 The other possible symmetry is SO(3). It turns out that the solution is really similar to SO(2) for $N=3$. SO(3) has three independent generators:
 \begin{equation}
  \begin{pmatrix} 0 & 1 & 0 & 0 \\ -1 & 0 & 0 & 0 \\ 0 & 0 & 0 & 0 \\ 0 & 0 & 0 & 0 \end{pmatrix},  \begin{pmatrix} 0 & 0 & 1 & 0 \\ 0 & 0 & 0 & 0 \\ -1 & 0 & 0 & 0\\ 0 & 0 & 0 & 0 \end{pmatrix},  \begin{pmatrix} 0 & 0 & 0 & 0 \\ 0 & 0 & 1 & 0 \\ 0 & -1 & 0 & 0 \\ 0 & 0 & 0 & 0 \end{pmatrix}.
 \end{equation}
 \eqref{eq:N=3:T_req} now gives the following nonzero components (up to permutations of the indices):
 \begin{align}
 \begin{split}
&  Q_{114}=Q_{224}=Q_{334}\equiv x,\\
  & Q_{444} \equiv y
  \label{eq:allQeq}
 \end{split}
 \end{align}
 with 
all the other terms being zero. 
The critical points now have the topology of a 2-sphere $S^2$, $\phi_1^2+\phi_2^2+\phi_3^2=r_0^2$,
obtained by a straightforward extension of \eq{eq:N=3:crit_point_S1}. 
The quantity \eqref{eq:defofpsip} can also be reduced to a single integral:
 \begin{align}
  \Psi(Q) &= \int_{\mathbb{R}^4} d\phi \text{\ } e^{i \left(\phi_1^2+\phi_2^2+\phi_3^2+\phi_4^2 + x (\phi_1^2 + \phi_2^2 + \phi_3^2)\phi_4 + y \phi_4^3\right)-\epsilon \phi^2},\nonumber\\
   &= \pi \sqrt{\pi} e^{i 3\pi/4} \int_\mathbb{R} d\phi_4 \frac{e^{i(\phi_4^2 + y \phi_4^3) - \epsilon \phi_4^2}}{(1+x \phi_4+i \epsilon)^{3/2}},
 \end{align}
 where the only relevant difference from the $N=3$ case, \eq{eq:N=3:Psiintegral}, is the exponent of the denominator.
One can also evaluate certain asymmetric configurations by
generalizing \eq{eq:allQeq} to $Q_{114}\neq Q_{224} \neq Q_{334} \neq Q_{114}$
parameterized by $x_{1,2,3}$, respectively.
This will again give just a single integral which is reasonably computable numerically:
 \begin{align}
  \Psi(Q) = &\pi \sqrt{\pi} e^{i 3\pi/4} \int_\mathbb{R} d\phi_4\ e^{i(\phi_4^2 + y \phi_4^3)-\epsilon \phi_4^2} \nonumber \\
  &\cdot \frac{1}{\sqrt{1+x_1 \phi_4 + i \epsilon}\sqrt{1+x_2 \phi_4 + i \epsilon}\sqrt{1+x_3 \phi_4 + i \epsilon}}.
  \label{eq:Neq4SO3}
 \end{align}
 Again, setting $x_1=x_2=x_3$ in this parametrization gives the $SO(3)$ symmetric configurations. 
 Numerical calculations show that this is indeed the preferred configuration, as given in Fig.\ref{fig:N=4SO(3)}. 
 In the figure, one can also observe lower peaks emanating in three directions, $z_1=0,z_2=0,$ and $z_1=z_2$, respectively, 
 which are the $SO(2)$ symmetric configurations. 
 
 The $N=4$ case demonstrates an interesting aspect of the mechanism explained in section \ref{sec:mechanism}. Though the critical points of $SO(3)$ are expected to have a larger individual contribution to $\Psi$ (in other words; a higher peak) because the space of the critical points is larger, the configuration space with $SO(2)$ symmetry is always larger. 
 This interplay between the size of the critical points space and the size of the configuration space with a certain symmetry is expected to be important in the large scale behavior of, for instance, the canonical tensor model.
 
\section{Summary and future prospects}
\label{sec:summary}
An important unresolved question in quantum gravity is 
how a macroscopic spacetime like the universe
can 
emerge from a, probably chaotic, microscopic theory. A few recent developments suggest the importance of Lorentz treatment in such emergence 
\cite{Ambjorn:2004qm,VanRaamsdonk:2010pw,Feldbrugge:2017kzv}. 
In view of this, we have proposed a simple mechanism which highlights partial symmetries 
based on collective quantum coherence. 
Starting from a widely used setup with underlying symmetries, we have shown that configurations 
symmetric under part of the underlying symmetries are 
highlighted by collective quantum coherence over generic non-symmetric configurations. 
Qualitative arguments show that
configurations invariant under partial symmetries with large dimensional representations are preferred.
This seems to be in accord with the present form of the universe: relatively small dimensional symmetries
like the Poincare/de Sitter symmetries represented on the whole universe.
To demonstrate this mechanism, we have shown the occurrence of the phenomenon in some simple computable examples
in the randomly connected tensor network \cite{Sasakura:2014zwa,Sasakura:2014yoa,RevModPhys.80.1275}, 
which is closely related to the canonical tensor model \cite{Sasakura:2011sq,Sasakura:2012fb}, 
a model of quantum gravity. 
We have observed strikingly sharp highlighting of configurations invariant under
either Abelian or non-Abelian symmetries and also observed 
the phenomenon of charge quantization even in the Abelian case.
Whether macroscopic spacetimes can be generated through the mechanism 
in the canonical tensor model will be an immediate future study.

We have put our mechanism in a familiar form, treating it in a rather general way. However, this does not imply that it is directly applicable for just any theory. Formally this could for instance be placed in a naive quantum gravity context by considering the metric at a certain time-slice as the configuration (called $Q$ in Section~\ref{sec:mechanism}) and the histories of the universe as integration variables (called $\phi$ in Section~\ref{sec:mechanism}), where the diffeomorphism invariance is the underlying symmetry. 
However, the classical equation of motion corresponding to the critical equation \eqref{eq:criticalpoints} cannot be solved in a fully consistent manner for the Minkowski signature, 
because solutions will generally encounter singularities at certain spacetime locations. 
This makes the stationary phase analysis unreliable, and one would have to treat the system in a
well-defined quantum mechanical manner from the beginning.
In fact, there have been some recent developments in the asymptotic safety program~\cite{Reuter:2012id, Eichhorn:2017egq}, which 
studies gravity as a non-perturbative quantum field theory. 
The program is usually studied in a Euclidean context, but can often be Wick rotated to Lorentzian signature~\cite{Manrique:2011jc}. 
Considering technical difficulties immediate results would not be expected, 
but there would be possibilities for our effect to appear in this context.   

Applying the mechanism to CDT also seems to be a non-trivial exercise, as the symmetries of the Regge action are usually not that simple, and after performing a Wick rotation the weights of the partition function are real instead of imaginary. The causality condition, restricting the triangulations to be compatible with
a space-time decomposition, 
makes this Wick rotation rigorous and is considered to be tightly related to the success of CDT. 
The necessity of such restriction may be compared to what one would find after
some analytic continuation of
the integral \eq{eq:psi} to make the weights real.\footnote{The actual process of performing such analytic continuation would 
be given by the Picard-Lefschetz theory. According to this,
an integral like \eq{eq:psi} can be expressed as a summation of components, each of which is 
a product of an essentially real integral and an overall complex factor.} 
According to the mechanism, one would expect 
that symmetric configurations would obtain large weights, while non-symmetric configurations would be suppressed.    
It would also be nice to look into the Regge calculus from a Lorentzian viewpoint: 
the symmetries of the classical solutions\footnote{In our setup these are called critical points.} are enhanced
for flat geometries or geometries with a small positive cosmological constant~\cite{Bahr:2009ku}. 
The mechanism would suggest the preference of such geometries with small curvatures.
  
Though the mechanism might not be always directly applicable in its current form, the mechanism does seem to be more reliable if the configuration and integration variables ($\phi$ and $Q$) are more independent, so that the form of \eqref{eq:psi} is directly satisfied. This could for instance be a statistical system with a sufficient amount of variables or a quantum mechanical wave function with an integral representation which has a similar form as \eqref{eq:psi}. This is why the mechanism works well in the case of the random tensor network discussed in Section~\ref{sec:model} , 
and we expect the mechanism to work well for the canonical tensor model.

An important characteristic of our mechanism is that peaks can appear only when there exist collective coherent phenomena along orbits of critical points. Therefore, a large peak inevitably contains a large number of degrees of freedom contributing to form it. Due to this collective nature, statistical tools could be useful to further analyse this mechanism, and characterise these preferred configurations. This might give a connection to the thermodynamic description of 
spacetime discussed in the literatures \cite{Bekenstein:1973ur,Jacobson:1995ab,Padmanabhan:2009vy,Verlinde:2010hp}.

\vspace{1cm}
\section*{Acknowledgements}
The work of N.S. is supported in part by JSPS KAKENHI Grant Number 15K05050. 
The work of D.O. is supported in part by the Hendrik Mullerfonds.
N.S. would like to thank Y.~Sato and B.~Fraser for some initial discussions during his stay in 
Chulalongkorn University. We would also like to thank S.~Hirano and H.~Nielsen for some intriguing discussions
in the occasions of their visits to Yukawa Institute for Theoretical Physics. 

\appendix

\section{Detailed calculations for $N=3$}
The contribution of the isolated critical points given in \eqref{eq:N=3:crit_point_1} and \eqref{eq:N=3:crit_point_2} can be obtained by perturbing $S(\phi^\sigma + \delta \phi)$ up to second order and performing Gaussian integration:
  \begin{align}
   \Psi(x,y)_{\text{i}} &\approx \sum_{\sigma:{\rm isolated}} \int_{\mathbb{R}^3} 
   d\delta \phi \ e^{i S(\phi^\sigma, x,y) + \frac{i}{2} S_{ab}^\sigma \delta \phi^a \delta \phi^b}  \nonumber\\
    &= \sum_{\sigma:{\rm isolated}} e^{i S(\phi^\sigma, x,y)}\prod_{l=1}^3  \left(\frac{2\pi i}{\alpha^\sigma_l}\right)^{\frac{1}{2}}\nonumber\\
    &= \pi^{3/2} e^{i 3 \pi/4} \left(1 + i\left(\frac{3y}{2x - 3y}\right) e^{i \frac{4}{27 y^2}}\right).
  \end{align}
  Here $S_{ab}^{\sigma} = \left.\frac{\partial^2 S}{\partial \phi^a \partial \phi^b}\right\vert_{\phi=\phi^\sigma}$,
  and $\alpha^\sigma_l$ are the eigenvalues of the matrix $S^\sigma_{ab}$. For the evaluation of the square root we choose the principal square root,  taking the branch cut along the whole negative real axis. 
  
  Next we evaluate the contribution from the continuous set of critical points \eqref{eq:N=3:crit_point_S1}. 
  Since these critical points are degenerate, it is convenient to integrate over the degeneracy
  first.
For this, polar coordinates, where $\phi_1 = r \cos(\theta), \phi_2 = r \sin(\theta)$,  are most convenient since the critical points are degenerate along the angular direction. Plugging these coordinates into \eqref{eq:N=3:S_symmetric} gives a trivial integration over $\theta$, removing the degeneracy. The remaining Gaussian integrals can be evaluated to be:
  \begin{align}
      \Psi(x,y)_{\text{c}} &
\approx 2\pi e^{i\frac{x-y}{x^3}} \int r d r\, d\delta \phi_3 \ e^{i ((1-\frac{3y}{x})\delta\phi_3^2 + 2 r_0 x \delta r \delta\phi_3)}\nonumber\\
      &= \frac{2\pi^{3/2}}{\sqrt{-i(1-3y/x)}} e^{i\frac{x-y}{x^3}}
       \int_0^\infty r d r \  e^{-i \frac{r_0^2 x^2 \delta r^2}{1-3y/x}}\nonumber\\
      &= \frac{2\pi^{3/2}}{\sqrt{-i(1-3y/x)}} e^{i\frac{x-y}{x^3}} 
      \int_{-r_0}^\infty (\delta r + r_0) d\delta r \text{\ } e^{-i \frac{r_0^2 x^2 \delta r^2}{1-3y/x}}\nonumber\\
      &= \frac{\pi^2}{|x|} e^{i\frac{x-y}{x^3}} \left(1 + \text{erf}\left( \sqrt{\frac{i x^2 r_0^4}{1-3y/x}}\right)\right)\nonumber\\
 	&\ \ \ \ + \pi^{3/2} 
	\frac{ \sqrt{-i(1-3y/x)}}{r_0^2 x^2} e^{-i \frac{r_0^4 x^2}{1-3y/x}}e^{i\frac{x-y}{x^3}}\nonumber \\
      & \xrightarrow[r_0\rightarrow\infty]{} \frac{2\pi^2}{|x|} e^{i \frac{x-y}{x^3}}\delta_{\text{sgn}(x), -\text{sgn}(y)},
      \label{eq:N=3:Psiapprox_limit}
  \end{align}
  where $\delta \phi_3=\phi_3-\phi_3^\sigma,\ \delta r=r-r_0$, and $\text{sgn}(x)$ denotes the sign of $x$.
  Here the standard formulas, $\int_{-\infty}^{\infty} dx e^{i \frac{a}{2} x^2 + i J x} = \left(\frac{2 \pi i}{a}\right)^{1/2} e^{- i \frac{J^2}{2 a}}$, 
  $\int_{-b}^\infty dx x e^{i \frac{a}{2} x^2} = \frac{i}{a} e^{i \frac{a}{2} b^2}$,
 and $\int_{-b}^\infty dx e^{- i a x^2} = \frac{\sqrt{\pi}}{2 \sqrt{i a}}\left(1 + \text{erf}(\sqrt{i a b^2})\right)$ have been used. 
 In the last step the $r_0\rightarrow \infty$ limit was taken in which we 
  expect the continuous critical point to become most relevant. 
  This limit can be realized by taking $x\rightarrow 0$ with $\text{sgn}(x) = -\text{sgn}(y)$,
as can be seen in \eqref{eq:N=3:crit_point_S1}.
  
  Because of the simplicity of the current model, it is also possible to do part of the calculation exact. The integral which needs to be solved is:
  \begin{equation}
      \Psi(x,y) = \int_{\mathbb{R}^3} d\phi \ e^{i (\phi_1^2 + \phi_2^2 + \phi_3^2 + x(\phi_1^2 + \phi_2^2)\phi_3 + y \phi_3^3) - \epsilon \phi^2},
  \end{equation}
  where the $-\epsilon \phi^2$ term was added to make sure the integral converges as mentioned in \eqref{eq:defofpsip}. The $\phi_1$ and $\phi_2$ integrals can be done readily since they are just Gaussian integrals, and using 
  $\int_{-\infty}^{\infty} dk e^{i a k^2} =  \sqrt{i\frac{\pi}{a}}$ one finds
  \begin{align}
      \Psi(x,y) &= i \pi \int_\mathbb{R} d\phi_3\ \frac{e^{i (\phi_3^2 + y \phi_3^3)-\epsilon \phi_3^2}}{1+x \phi_3 + i \epsilon}\nonumber\\
      &\equiv i \pi \int_{\mathbb{R}} d\phi_3 f(x,y,\phi_3) \label{eq:N=3:Psiintegral}.
  \end{align}
  Interestingly, the integrand has a simple pole at $\phi_3 = -\frac{1+i \epsilon}{x}$. This value of $\phi_3$ indeed
  agrees with that of the continuous critical point found in \eqref{eq:N=3:crit_point_S1}. 
  This suggests that the contribution of the pole is related to the continuous critical point, as will be shown more explicitly below.   
    
\begin{figure}
  \centering
  \resizebox{0.5\textwidth}{!}{
  \begin{tikzpicture}[decoration={markings,
mark=at position 4cm with {\arrow[line width=1.5pt]{>}},
mark=at position 8.6cm with {\arrow[line width=1.5pt]{>}},
mark=at position 13.2cm with {\arrow[line width=1.5pt]{>}},
mark=at position 17.8cm with {\arrow[line width=1.5pt]{>}}
}
]
   
    \draw[thick,->] (-5,0) -- (5,0) coordinate (xaxis);
    \draw[thick,->] (0,-1) -- (0,2.5) coordinate (yaxis);

   \path[draw,line width=1pt,postaction=decorate] (-4,0) node[below] {$-L$} -- (4,0) node[below] {$L$} -- (4,1.2)  -- (-4,1.2) -- (-4,0);

   \node[below] at (xaxis) {$\text{Re}(\phi_3)$};
   \node[left] at (yaxis) {$\text{Im}(\phi_3)$};
   \node[below left] {$O$};
   \node at (0.2,1.6) {$\lambda$};
   \node at (-2,-0.25) {$\ast$};
   \node at (-2,-.5) {\small{$x>0$}};
   \node at (2,0.25) {$\ast$};
   \node at (2,.5) {\small{$x<0$}};
  \end{tikzpicture}
  }
  \caption{The contour $\gamma$ in \eq{eq:contourintegral}. 
  With this contour, one can change the integration contour of $\phi_3$ in \eq{eq:N=3:Psiintegral} to that in
   \eq{eq:N=3:Psiintegral_contour}, where
$\lambda$ must have the same sign as $y$ for the limit $L\rightarrow \infty$ and $\epsilon \rightarrow 0^+$.
 There is a pole at $\phi_3=-\frac{1+i\epsilon}{x}$ shown by $\ast$, 
 and only for $\text{sgn}(x)=-\text{sgn}(y)$ the pole will lie inside the contour.
 The orientation of the contour depends on the sign of $\lambda$, generating the factor of $\text{sgn}(\lambda)$
 in \eq{eq:contourintegral}.
}
  \label{fig:N=3:contour}
  \end{figure}
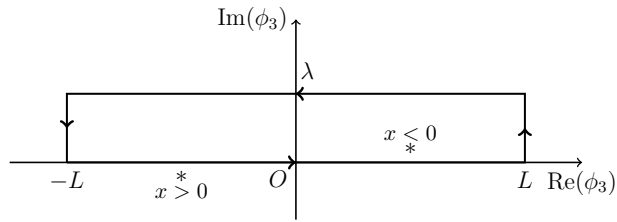

  The contribution of the pole of the integration of \eqref{eq:N=3:Psiintegral} can be picked out
   by taking the contour of Fig.\ref{fig:N=3:contour} in the complex $\phi_3$ plane.
Here, $\lambda$ is an arbitrary real number with the same sign as $y$, 
the reason of which will become apparent shortly. The residue theorem then gives:
  \begin{align}
      \int_\gamma d\phi_3 f(x,y,\phi_3) &= \text{sgn}(\lambda)\, 2 \pi i\, \text{Res}\left(f(x,y,\phi_3), -\frac{1}{x} \right)\nonumber\\
      &= -\frac{2\pi i}{|x|}e^{i\frac{x-y}{x^3}}\delta_{\text{sgn}(x), -\text{sgn}(y)}.
      \label{eq:contourintegral}
  \end{align}
  So, by taking $L\rightarrow \infty$ on account of $\text{sgn}(\lambda)$=$\text{sgn}(y)$,  the integral \eqref{eq:N=3:Psiintegral} can be rewritten:
  \begin{align}
  \begin{split}
      \Psi(x,y) = i \pi \int_{\mathbb{R}} d\phi_3\ f(x,y,\phi_3+i\lambda) \\
      + \frac{2\pi^2}{\vert x\vert} e^{i\frac{x-y}{x^3}}\delta_{\text{sgn}(x), -\text{sgn}(y)} \label{eq:N=3:Psiintegral_contour}.
  \end{split}
  \end{align}
Here note that there is no $\epsilon$ in the expression: 
we have been able to safely take the $\epsilon\rightarrow 0^+$ limit, because the integrand of the first term
damps exponentially $\sim \exp(-3 y \lambda \phi_3^2)$ and no regularization is necessary.
By comparing with \eqref{eq:N=3:Psiapprox_limit}, it can be seen that the continuous critical point 
is the source of the contribution of the pole for $r_0\rightarrow \infty$ (or $x\sim 0$).

\printbibliography

\end{document}